\documentclass[12pt,draftclsnofoot, onecolumn]{IEEEtran}
\usepackage{graphicx,cite,epsfig,amssymb,amsmath,url,stfloats,latexsym,threeparttable,balance}
\usepackage{multirow}
\usepackage{color,epstopdf}
\usepackage{booktabs}
\usepackage{setspace}
\usepackage{amsbsy}
\usepackage{cite}
\usepackage{mathrsfs}
\usepackage{amsfonts}
\usepackage{subfig}

\begin{document}
\title{\mbox{}\vspace{1.5cm}\\
\textsc{RIS-Based On-the-Air Semantic Communications - a Diffractional Deep Neural Network Approach} \vspace{1.5cm}}
\author{Shuyi Chen,~\IEEEmembership{Member,~IEEE}, Yingzhe Hui,~\IEEEmembership{
Graduate Student Member,~IEEE}, Yifan Qin, Yueyi Yuan, Weixiao Meng,~\IEEEmembership{Senior Member,~IEEE}, Xuewen Luo, and Hsiao-Hwa Chen{$^{^\dagger}$},~\IEEEmembership{Fellow,~IEEE}
\thanks{Shuyi Chen (e-mail: {\tt chenshuyitina@gmail.com}), Yingzhe Hui (e-mail: {\tt yingzhe\_hui@stu.hit.edu.cn}), Yueyi Yuan (e-mail: {\tt yuanyueyi@hit.edu.cn}), Weixiao Meng (e-mail: {\tt wxmeng@hit.edu.cn}) and Xuewen Luo (e-mail:{\tt luoxw@hit.edu.cn}) are with the Communications Research Center, Harbin Institute of Technology, China. Yifan Qin (e-mail: {\tt qinyifan@hrbeu.edu.cn}) is with the Ministry of Education Key Lab of In-fiber Integrated Optics, Harbin Engineering University, China. Hsiao-Hwa Chen (e-mail: {\tt hshwchen@ieee.org}) (the corresponding Author) is with the Department of Engineering Science, National Cheng Kung University, Taiwan.}
\thanks{This work was supported in part by the National Science Fund for Young Scholars No. 62201176, Young Elite Scientist Sponsorship Program by CAST No. YESS20210339, Fellowship of China Postdoctoral Science Foundation No. 2021TQ0092, Heilongjiang Postdoctoral Fund No. LBH-Z21001, and Taiwan Ministry of Science and Technology (109-2221-E-006-175-MY3 and 109-2221-E-006-182-MY3).}
}
\renewcommand{\baselinestretch}{1.2}
\thispagestyle{empty} \maketitle \thispagestyle{empty}
\setcounter{page}{1}

\begin{abstract}
Semantic communication has gained significant attention recently due to its advantages in achieving higher transmission efficiency by focusing on semantic information instead of bit-level information. However, current AI-based semantic communication methods require digital hardware for implementation. With the rapid advancement on reconfigurable intelligence surfaces (RISs), a new approach called on-the-air diffractional deep neural networks (D$^2$NN) can be utilized to enable semantic communications on the wave domain. This paper proposes a new paradigm of RIS-based on-the-air semantic communications, where the computational process occurs inherently as wireless signals pass through RISs. We present the system model and discuss the data and control flows of this scheme, followed by a performance analysis using image transmission as an example. In comparison to traditional hardware-based approaches, RIS-based semantic communications offer appealing features, such as light-speed computation, low computational power requirements, and the ability to handle multiple tasks simultaneously.
\end{abstract}

\IEEEpeerreviewmaketitle

\vspace{0.4in}
\section{INTRODUCTION}
5G networks have been utilized to facilitate a wide range of services on a single platform, ranging from enhanced Mobile BroadBand (eMBB) communications to virtual reality, automated driving, Internet of Things, etc. However, the challenge arises as more resources are required to accommodate upcoming new services in future wireless communication networks. The scarcity in radio spectrum inspires us to reconsider the three levels of communications (i.e., technical, semantic, and effectiveness) as identified by Warren Weaver in 1998, and to focus on constructing a semantics-level, instead of a bit-level, communication system.

Thanks to the advancements in artificial intelligence (AI), the field of semantic communications has gained significant attention recently. Current research in semantic communications can be broadly categorized into two types: data-reconstruction-oriented and task-oriented semantic communications~\cite{Zhang2022SemICC}. The former aims to reconstruct the source data, similar to traditional communications, but with the use of semantic features extracted at the receiver. On the other hand, the latter focuses on directly performing tasks using the semantic features. Task-oriented semantics, with their emphasis on the communication goal, enable us to extract the data that specifically conveys the intended information from the source.

Most studies on AI-based semantic communications have always assumed that AI is performed on digital hardware. Meanwhile, the emergence of application-specific integrated circuits (ASIC) for AI has significantly improved computation performance while reducing power consumption. This advancement makes semantic communication applicable in future wireless networks. However, in addition to hardware-based AI technologies, recent studies have also suggested using physical layer solutions with on-the-air or all wave-based neural networks, known as diffractional deep neural networks (D$^2$NN) ~\cite{Wright2022PNNNature,Meta2022CuiNature}. In this approach, the computation of a layer in hardware-based neural networks is equivalent to the transmission of the wave through a specific device. These methods leverage physical layer signal transformations directly, enabling more efficient computations compared to conventional hardware-based paradigms. This approach offers a scalable, energy-efficient, and faster machine learning implementation. Moreover, reconfigurable intelligence surfaces (RISs), which can adjust the amplitude and phase of transmitted or reflected waves, serve as a promising candidate for implementing D$^2$NN~\cite{Meta2022LuoLight}.

Inspired by recent advancements in RIS-based on-the-air D$^2$NN, we are interested in exploring the potential of enabling semantic communications on RISs. This allows the computational process involved in semantic communications to be carried out while transmitting wireless signals through RISs. The proposed scheme offers several attractive features when compared to traditional hardware-based approaches, including fast computation, low power requirements, and the ability to handle multiple tasks concurrently.

In summary, in this paper we aim to motivate a paradigm shift from the mainstream research on hardware-based semantic communications towards RIS-based on-the-air D$^2$NN semantic communications. The rest of the paper can be outlined as follows. First, we present some background knowledge on complex-valued convolutional neural networks (CvCNNs) and RIS-based transmission/computation, to help readers understand how RIS can realize on-the-air semantic communication. A system model of RIS-based semantic communication is presented next, consisting of transceiver structure, data- and control-flows, and an example of image transmission. Then, the advantages and disadvantages of the proposed scheme are elaborated, followed by the conclusions.

\vspace{0.4in}
\section{BACKGROUND KNOWLEDGE}
As RIS-based semantic communication operates directly on wireless signals, the usage of real-number convolutional neural network (CNN) is not applicable. Instead, we need to employ complex-valued CNN (CvCNN) to design the neural networks. Therefore, this section reviews some pertinent information about CvCNN. Additionally, since on-the-air signal processing on RISs is a novel research area, we will also showcase some relevant studies to demonstrate the feasibility of RIS-based semantic communications.

\subsection{CvCNN}
CNN has been widely used in semantic communications. Although there have been few studies comparing the performance of CvCNN and CNN in semantic communications, many studies have demonstrated the advantages of using CvCNN in image and signal processing. While CvCNN requires more complex computations and longer computation time, it can leverage the benefits of complex representations and has the potential to facilitate easier optimization, better generalization, faster learning, and the ability to achieve noise-robust memory mechanisms \cite{Trab2018CompICLR}. The main difference between CvCNN and CNN lies in the fact that CvCNN uses complex numbers to represent its input, weights, and biases. As a result, complex multiplications, complex convolutions, and complex activation functions are required.

\begin{figure}
\centering
\includegraphics[width=16cm]{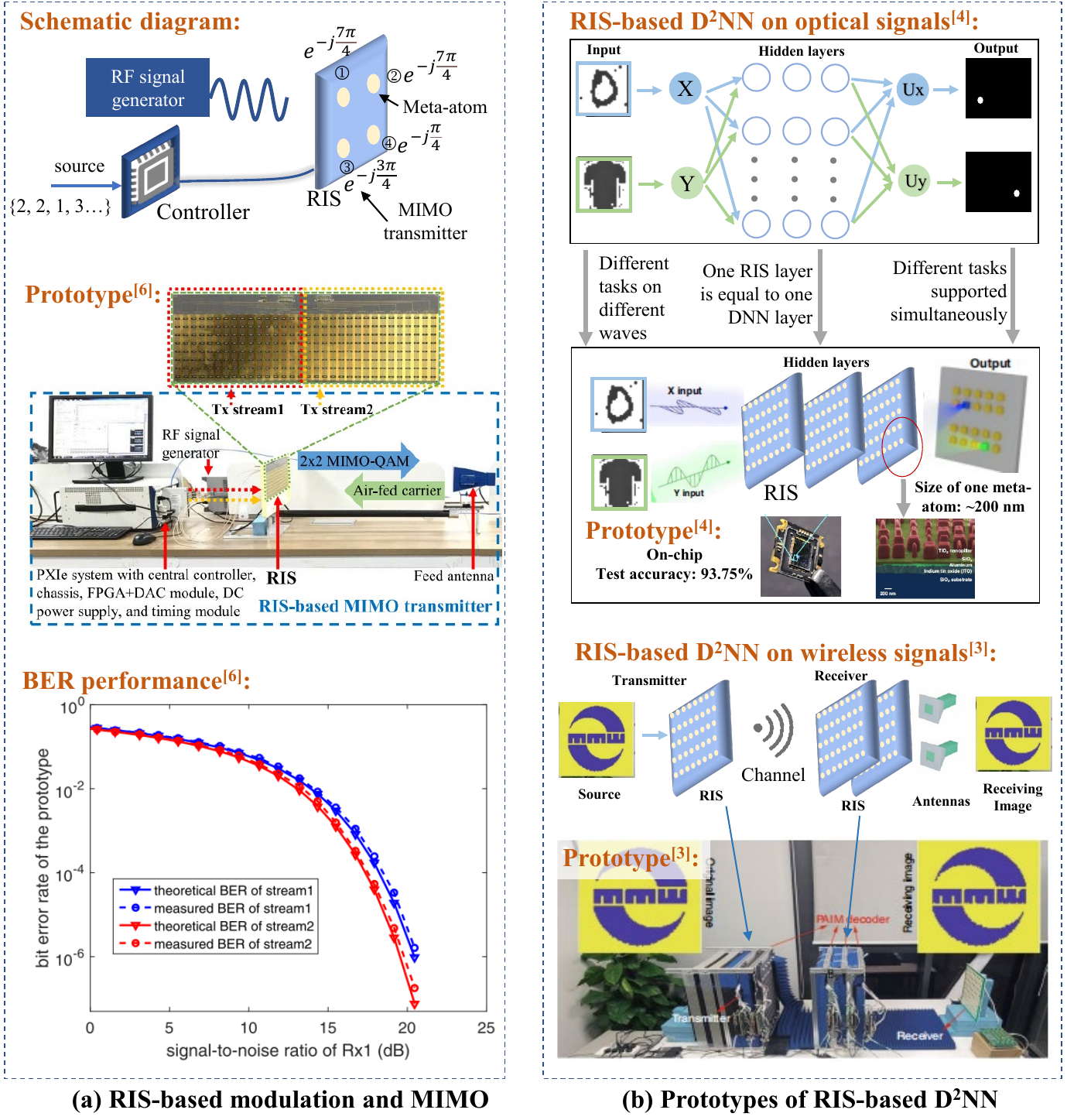}
\caption{\small{Background knowledge of RIS-based semantic communications, where (a) demonstrates RIS-based modulation/MIMO~\cite{Tang2020RISJSAC}, and (b) shows two prototypes of RIS-based D$^2$NN. The first RIS-based D$^2$NN prototype operates on optical signals, and shows that RISs can support multiple tasks simultaneously. The second one operates on wireless signals, which illustrates that RISs help to construct an end-to-end communication system. }}
\label{background}
\end{figure}

\subsection{RIS-based Modulation and MIMO}
RIS is a programmable metasurface made up of tiny meta-atoms that are regularly arranged. Each programmable meta-atom usually consists of a metal pattern, dielectric, and tunable component. By adjusting the electrical parameters of the tunable component through an external control signal, the response of each meta-atom, including phase and amplitude, can be modified. This allows RIS to function as a transmit antenna with programmable electromagnetic properties. Additionally, the RIS-based wireless transmitter can directly modulate carrier signals without using traditional RF chains~\cite{Tang2020RISJSAC}. The prototype and BER performance of RIS-MIMO are depicted in Fig.~\ref{background}(a)~\cite{Tang2020RISJSAC}. To provide a clearer understanding of RIS-based modulation and MIMO, a simple example is shown in the upper part of Fig.~\ref{background}(a), where four meta-atoms exist and the RIS acts as a transmitter for a 4-antenna MIMO system. When the source array is \{2,2,1,3...\}, the controller adjusts the responses of the meta-atoms based on the first four symbols. This results in the output signals from the four meta-atoms being \{$e^{-j\frac{7\pi}{4}},e^{-j\frac{7\pi}{4}},e^{-j\frac{3\pi}{4}},e^{-j\frac{1\pi}{4}}$\}, respectively. The subsequent four symbols are then used to modify the response of the RIS, achieving both modulation and MIMO transmission. It should be noted that while RIS-based transmitters show potential for ultra-massive-MIMO and holographic MIMO, there are still hardware constraints, such as phase-dependent amplitude and discrete phase shift, that limit their application scenarios and make it more challenging to implement high-order modulation using RIS-based transmitters.


\subsection{RIS-based AI}
On-the-air AI on RISs is an emerging research area. While no prototype of a RIS-based CvCNN has been implemented, it has been demonstrated that various layers of a CvCNN, i.e., fully-connected, convolution and activation layers, can be achieved by RISs, which can be explained as follows.
\begin{itemize}
\item \textbf{Fully-connected layer:} An on-chip prototype of a D$^2$NN with fully-connected layers has been established in~\cite{Meta2022LuoLight}. As shown in the upper part of Fig.~\ref{background}(b), two distinct functions, i.e., object recognition and hand-written number recognition, are supported by the same set of RISs simultaneously. The vertical-polarized optical wave is responsible for object recognition, while the horizontal-polarized optical wave handles the other task of hand-written number recognition. Each RIS layer corresponds to a hidden layer, and the signal diffraction between two RIS layers enables full connection between the two layers. The optical input signal carries the information of the object and hand-written number. As the signal passes through the RIS layers, which correspond to the hidden layers in a conventional DNN, two different recognition functions are achieved. As shown in Fig.~\ref{background}(b), the TiO$_2$ metasurface is extremely small, with each meta-atom measuring about 200 nm in size. Meanwhile, other materials, including quartz, polymethyl methacrylate, gold, aluminum, hydrogen silsesquioxane and SiO$_x$, are used and seven major procedures are required to fabricate this on-chip prototype~\cite{Meta2022LuoLight}. In~\cite{Meta2022CuiNature}, the functions of fully-connected layers are achieved on wireless signals such that the images are successfully transmitted, as shown in the lower part of Fig.~\ref{background}(b). Unlike the prototype in~\cite{Meta2022LuoLight}, the response of each meta-atom can be controlled in real time as an integrated circuite module is attached to each meta-atom. Moreover, an amplifier device is also included in each meta-atom to realize different gain levels in programmable way~\cite{Meta2022CuiNature}. 
\item \textbf{Convolution layer:} A prototype of a complex convolution layer on the RIS is provided in~\cite{Fu2022ConvLight}, where the authors used the prototype to realize the function of image feature extraction.
\item\textbf{Activation function:} As to the complex-valued activation functions, it has been proved theoretically that RISs can support the nonlinear functions either by making the amplifier work in a nonlinear range or integrating the phase modulation in each meta-atom~\cite{Meta2022CuiNature}.
\end{itemize}

\section{RIS-BASED ON-THE-AIR SEMANTIC COMMUNICATIONS}
\begin{figure}
\centering
\includegraphics[width=16cm]{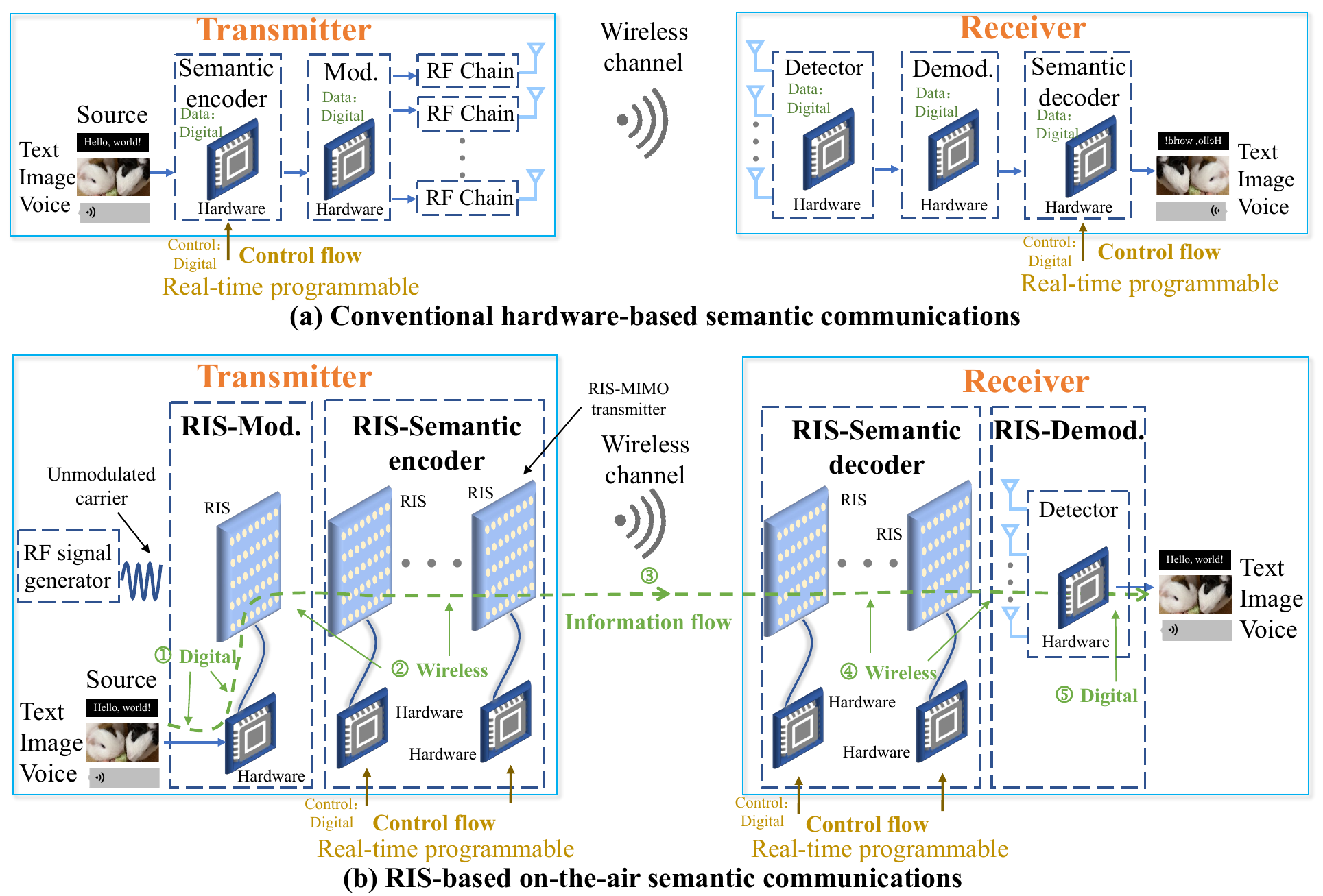}
\caption{\small{Comparison between traditional semantic communications and RIS-based on-the-air semantic communications, where "mod." and "demod." are short for modulation and demodulation, respectively. }}
\label{systemmodel}
\end{figure}

Various kinds of semantic communications have been suggested. In this paper, we adopt the most commonly used version, which considers semantic communication as a solution based on machine learning for joint source and channel coding. The block diagram of a traditional point-to-point semantic communication is depicted in Fig.~\ref{systemmodel}(a), where semantic encoding is carried out directly on the digital source using hardware, and then the encoded digital data is modulated and converted into an RF signal.

In this paper, we propose an on-the-air computation approach based on RISs, as illustrated in Fig.~\ref{systemmodel}(b). The semantic coding is directly applied to RF signals using CvCNN. In this section, we will provide a detailed explanation of the system model for our proposed scheme, starting with the description of the transceiver's structure. Additionally, since the data and control flows are completely separated in our proposed scheme, we will analyze these two types of flows separately. Finally, we will demonstrate the performance of the proposed scheme using an example of image transmission.

\subsection{Transceiver Structure}
As shown in Fig.~\ref{systemmodel}(b), the transmitter consists of a RIS-modulator and a RIS-semantic encoder, and the receiver requires symmetrical demodulator and decoder modules. The RIS-modulator is comprised of a single layer of RIS and a hardware controller, such as an FPGA, to adjust the response of each meta-atom based on input digital data. Prior to the RIS-modulator, an RF signal generator is utilized to generate RF carrier signals without any information. When these RF carrier signals pass through the RIS-modulator, the output amplitude and phase of each meta-atom are determined by the product of the incident electric field and the complex-valued transmission coefficient of the meta-atom. Thus, modulation and MIMO transmission are achieved through signal processing. The RIS-demodulator at the receiver employs antenna arrays to collect the electronic wave and achieve spatial diversity gain of the signals.

Then, let us focus on the RIS semantic encoder and decoder modules. Each module consists of a series of RISs and the corresponding controlling hardware. Each meta-atom functions as a neuron, and each RIS layer corresponds to a CvCNN layer. As we discussed in the previous section, multi-layer RISs can perform three functions of CvCNN, namely full connection, complex convolution, or complex-valued activation, to fulfill the fundamental requirements of semantic communications. Additionally, a training process similar to hardware-based semantic communications is required, but with a distinction that no data passes through the hardware in the RIS-semantic encoder.  Instead, the hardware is solely utilized to adjust the response of the meta-atom. This achieves a separation of data and control flows, resulting in an expected enhancement in security within the proposed scheme.

\begin{figure}
\centering
\includegraphics[width=9cm]{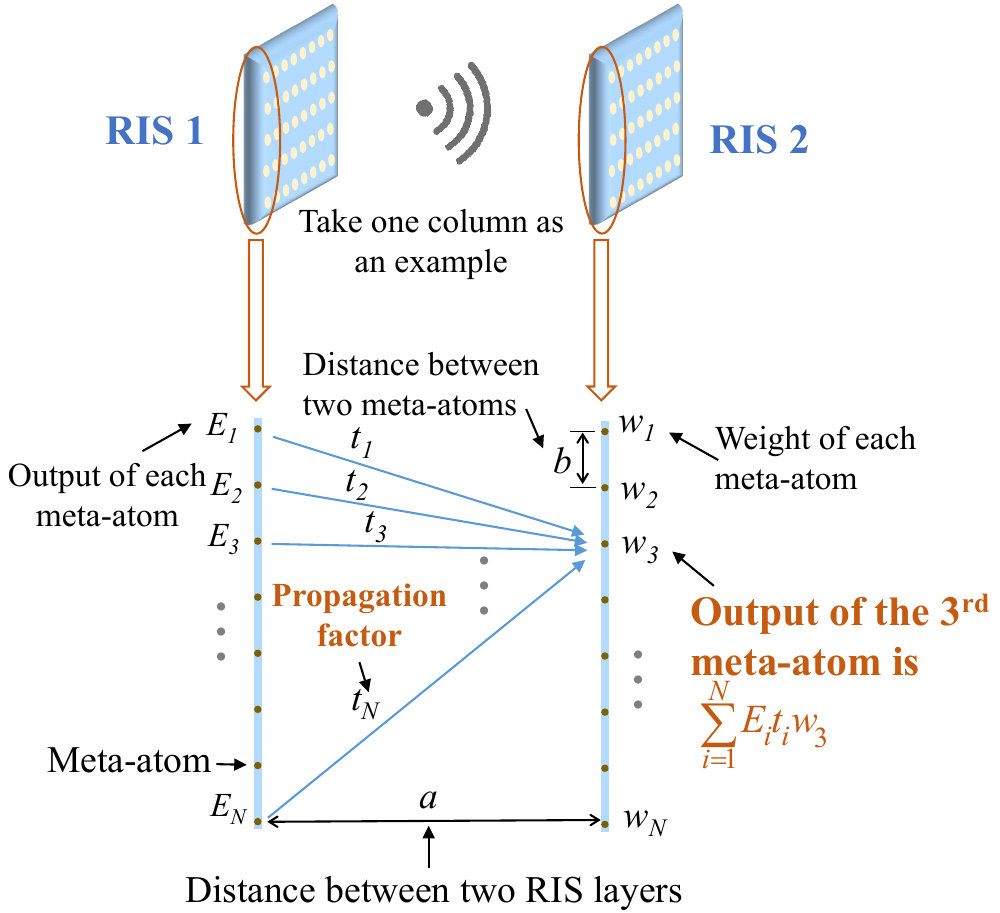}
\caption{\small{Signal propagation between two RIS layers. }}
\label{propagation}
\end{figure}

Meanwhile, it is important to highlight several distinctive characteristics of RISs that should be taken into consideration when determining the parameters of CvCNN. The propagation between two RIS layers is illustrated in Fig.~\ref{propagation}. The output of the second layer is influenced by both the input of the second layer and the weight of the second layer, mirroring the approach of a hardware-based CvCNN. However, it is worth noting that the input of the second layer is not identical to the output of the first layer. Due to the signal propagation between two RIS layers, an additional phase shift is expected, which is called propagation factor~\cite{Meta2022CuiNature}. Thus, in order to train a RIS-based CvCNN on hardware, it is important to include the propagation factor. As shown in the output of each meta-atom in Figure~\ref{propagation}, the impact of signal propagation between two RIS layers can be seen as adding an extra linear layer with fixed parameters between these two layers. The value of the propagation factor, obtained through the Rayleigh-Sommerfeld diffraction integral, depends on the distance between the two layers as well as the distance between two meta-atoms. Therefore, we have the ability to adjust the propagation factor as needed by modifying the distance between two RIS layers~\cite{Meta2022LuoLight}.

\subsection{Data- and Control-flows}
In the proposed scheme, the data and control flows are completely separated, as indicated by the green and yellow lines in Fig.~\ref{systemmodel}. The data flow can be roughly divided into five stages, represented by circled numbers. We will provide a detailed explanation of each stage below.
\begin{itemize}
\item The first stage involves mapping the digital source to a wireless signal using the RIS-modulator. Specifically, various types of sources such as texts, voices, and images are converted into digital data. The controller of the RIS-modulator then adjusts the responses of meta-atoms based on the inputs. For instance, when transmitting images, each pixel can be encoded directly on the output phase or the amplitude of each meta-atom. However, for other types of information, such as text or voice, real-number graphical expressions or complex-valued expressions are required to represent the data~\cite{Weng2021SemJSAC}. Since this paper focuses on the proposed scheme and not on the complex-value expression of different sources, images are chosen as the source in order to provide readers with a clearer understanding.
\item The data in the second to fourth stages is transmitted as wireless signals. In the second stage, the RIS-semantic encoder performs on-the-air computation on the wireless signals to achieve the functions of a CvCNN. The last RIS layer acts as transmit antennas. After passing through the wireless channels, which is referred to as the third stage, the data is collected by the RIS-semantic decoder at the receiver. Then, similarly, on-the-air computation is performed on the wireless signals to achieve the functions of CvCNN. Additionally, it should be noted that the dimension of the output layer of the RIS-semantic encoder is generally smaller than the input layer, as the purpose of semantic communications is to reduce the amount of transmitted data. However, depending on the specific tasks of semantic communication, the size of the output layer of the RIS-semantic decoder can be made equal to the layer of the RIS-modulator or smaller than the output layer of the RIS-semantic encoder.
\item The final step involves converting the RF signal into digital data. It is important to note that due to the limited computational capabilities of RISs, certain decision layers, such as the softmax layer in the RIS-semantic decoder, can be shifted to this stage. Therefore, digital computation can complement wireless signal computation, expanding the potential application scenarios of the proposed scheme.
\end{itemize}

The control-flow is represented by the yellow lines in Fig.~\ref{systemmodel}, and it can be distinguished from the data-flow. The assignment and real-time updating of the parameters follow the same procedure as the conventional semantic coding, which will not be discussed in detail in this paper due to space constraints. Here, we aim to highlight the control process of RISs' responses. Since each RIS layer is equipped with an FPGA, the FPGA arrays have the ability to generate various levels of bias voltages that can be applied to the integrated circuit of each meta-atom. This modulation enables the individual modulation of the transmission coefficient of each meta-atom~\cite{Meta2022CuiNature}. 

\subsection{Image Transmission as an Example}
As mentioned earlier, semantic communications can be divided into two categories: data-reconstruction-oriented and task-oriented schemes. In this section, we will focus on the data-reconstruction-oriented semantic communication to demonstrate the performance of the proposed scheme. We believe that if the proposed scheme can successfully perform data-reconstruction-oriented semantic communication, it can be adapted to support other types of semantic communications as well. For comparison, we will use a baseline scheme called joint source-channel coding (JSCC), described in~\cite{Bour2019JSSCTCCN}. The same dataset, CIFAR-10 image dataset, will be used to train and evaluate the proposed scheme in this paper. The proposed scheme was trained on the CIFAR-10 dataset with a specific signal-to-noise ratio (SNR), denoted as $SNR_{train}$, and then evaluated with varying SNR values, denoted as $SNR_{test}$. The training dataset includes 50,000 32 $\times$ 32 training images, while the testing dataset contains 10,000 images that are separate from the training set. We use a batch size of 128 samples and train our model for approximately 400 epochs, with the learning rate adjusted three times. By adjusting the values of $SNR_{train}$ and $SNR_{test}$, we can assess the performance of the proposed scheme under different channel conditions than what the system was initially designed for. This allows us to observe how well the proposed scheme handles changes in channel quality~\cite{Bour2019JSSCTCCN}.

\begin{figure}
\centering
\includegraphics[width=16cm]{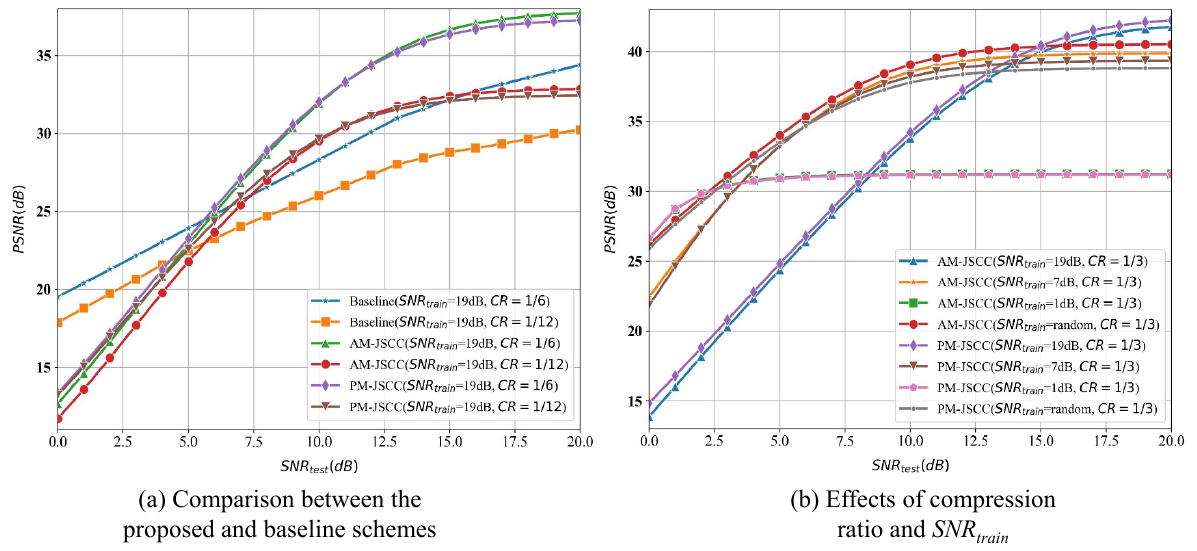}
\caption{\small{PSNR of the proposed scheme on CIFAR-10 test images with respect to SNR over an AWGN channel. (a) compares the proposed two schemes, i.e., AM-JSCC and PM-JSCC, with the baseline scheme from~\cite{Bour2019JSSCTCCN}, where $SNR_{train}$ is 19 dB and compression ratio (CR) is 1/6 or 1/12. (b) shows the effect of $SNR_{train}$. }}
\label{PSNR}
\end{figure}

Meanwhile, even though the date take complex values in the proposed scheme, only one dimension of the signal is used to modulate the data. In other words, to present a fair comparison to the real-valued CNN in the baseline scheme in~\cite{Bour2019JSSCTCCN}, the data was modulated either on the phase or the amplitude of the signals, and the former scheme is name as phase-modulated JSCC (PM-JSCC) and the latter is amplitude-modulated JSCC (AM-JSCC). The performance is measured using peak signal to noise ratio (PSNR), which calculates the ratio between the maximum signal power and the power of noise that affects the signal, as depicted in Fig.~\ref{PSNR}. Meanwhile, various compression ratios (CR) are configured, where CR indicates the amount of transmitted data in semantic communications relative to conventional communications. A lower CR implies a smaller amount of data is required for transmission, which results in a higher communication efficiency.

First, we will compare the PSNR performances of the proposed scheme and the baseline scheme from~\cite{Bour2019JSSCTCCN}, as shown in Fig.~\ref{PSNR}(a). It is evident that in most cases, the proposed AM-JSCC and PM-JSCC schemes outperform the baseline scheme. The primary reason is that the parameters in the proposed neural network are in the form of complex values. Despite requiring approximately four times as many floating-point operations, CvCNN surpasses CNN in terms of image reconstruction~\cite{Trab2018CompICLR}. Moreover, since the proposed scheme operates directly on the wireless signal, there is no additional computation cost associated with CvCNN. As a result, the proposed scheme achieves the same performance improvement as conventional CvCNN without incurring any extra computational overhead. The baseline scheme demonstrates improved robustness only when the channel conditions are very poor, such as when the signal-to-noise ratio (SNR) is below 3 dB. Similar performance can be observed for the AM-JSCC and PM-JSCC schemes, as they only differ in the modulation of the input signal. Therefore, we can analyze the resilience of the proposed JSCC scheme against variations in channel conditions, as depicted in Fig.~\ref{PSNR}(b). It is worth noting that when the SNR value used for training is set to a moderate or random value (i.e., the SNR is randomly generated to simulate the real channel environment), the proposed scheme exhibits greater robustness to fluctuations in channel quality. This means that even with a decrease in $SNR_{test}$, the decoder of the proposed scheme can still successfully reconstruct the original image.

\begin{figure}
\centering
\includegraphics[width=16cm]{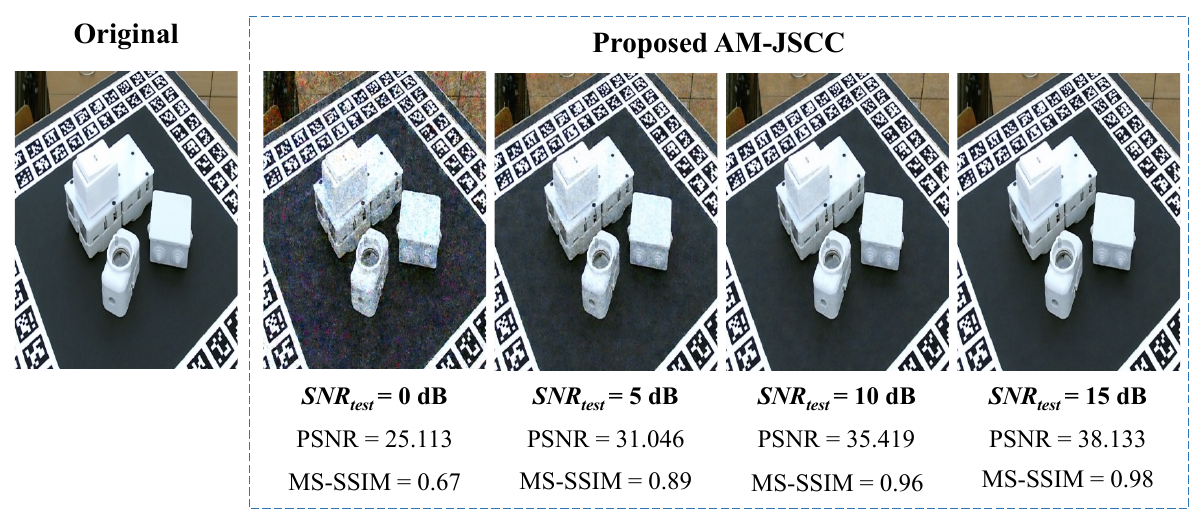}
\caption{Examples of reconstructed images produced by the proposed AM-JSCC scheme on T-LESS test images, where $SNR_{train}$ is 19 dB and CR is 1/3. }
\label{Reconstruction}
\end{figure}
Then, examples of reconstructed images based on the T-LESS dataset are shown in Fig.~\ref{Reconstruction}. This dataset consists of high-resolution images of industry-relevant objects that lack significant texture and do not have distinctive color or reflectance properties. In addition to PSNR, we also utilize the multi-scale structural similarity index (MS-SSIM), which is a perception-based model, to measure the quality of the reconstruction. It can be observed that even when the SNR is 0 dB, the image can still be reconstructed, and the reconstruction performance improves as the SNR increases. 

\section{PORS AND CONS OF RIS-BASED SEMANTIC COMMUNICATIONS }
There are several advantages to implementing RIS-based semantic communications, and some of the most appealing benefits are listed below.
\begin{itemize}
\item \textbf{Light-speed computation:} One of the most appealing features of the proposed plan is its fast computation speed. The need for ultra-low latency in future wireless communications puts a lot of pressure on data processing. Semantic communications have emerged as a way to reduce transmission time by using semantic information instead of actual source data. However, the involvement of AI will unavoidably increase the computation delay. We have summarized some commonly used CNN structures in research and provided the corresponding computation complexity on digital hardware in Table~\ref{Table}. Additionally, we have included the computation time of the proposed RIS-based CvCNN structures. It is evident that the computation time cannot be disregarded in services that require low delay. As the number of neurons in each layer increases, the computation time also increases significantly. However, in the proposed plan, the computation time is negligible because the total computation load of one layer is equal to the transmission process through one RIS, allowing for fast computation. Furthermore, the computation time in RIS-based semantic communications is not affected by the number of neurons in each layer. Therefore, the proposed scheme is more suitable for handling complex tasks. The estimated computation time of the proposed plan is also provided in Table~\ref{Table}. Since the distance between two RISs depends on the transmission frequency, we compare the computation time under millimeter and THz waves, respectively. Since the distance between two RISs can be adjusted based on different tasks, we assume an average of half wavelength for simplicity.

\begin{table*}[!t]
\scriptsize
\renewcommand{\arraystretch}{1.1}
\caption{Computation times of traditional and RIS-based semantic communications.}
\label{Table}
\centering
\begin{tabular}{|p{1.2cm}|p{4.4cm}|p{2.6cm} |p{2.6cm}|p{3cm}|}
\hline
\multirow{3}{*}{Scheme} &\multicolumn{2}{|c|}{Traditional}&\multicolumn{2}{|c|}{RIS-based}\\
&\multicolumn{2}{|c|}{semantic communications}&\multicolumn{2}{|c|}{semantic communications}\\
\cline{2-5}
&\centering{Parameters} &\centering{Computation time }& \centering{mmWave} (25 GHz)&THz (400 GHz)\\
\hline
\cite{Bour2019JSSCTCCN}& Eight 2.10 GHz Intel Xeon E5-2620V4 CPUs and a Tesla K80 GPU, to transmit an image of 768 $\times$ 512 pixels&387 ms on CPU, 18 ms on GPU&0.4 ns&0.025 ns\\
\hline
\cite{Wang2022JSSCJSAC}&Intel Xeon Gold 6226R CPU and a RTX 3090 GPU, to encode a frame of 1920 $\times$ 1080 resolution &280 ms&0.3 ns&0.019 ns \\
\hline
AM-JSCC&RTX 2080Ti GPU, to encode and decode a image of 32 $\times$ 32 pixels&Encode: 86 ms;
Decode: 79 ms&Encode: 0.2 ns;  Decode: 0.2 ns &Encode: 0.013 ns; Decode: 0.013 ns \\
\hline
PM-JSCC&RTX 2080Ti GPU, to encode and decode a image of 32 $\times$ 32 pixels&Encode: 81 ms;
Decode: 76 ms&Encode: 0.2 ns; Decode: 0.2 ns &Encode: 0.013 ns; Decode: 0.013 ns\\
\hline
\end{tabular}
\end{table*}

\item \textbf{Support different tasks simultaneously:} Owing to the polarization feature in RIS, orthogonal electromagnetic waves can pass through or be reflected by the RIS at the same time, carrying signals for the same or different tasks. In \cite{Meta2022CuiNature}, only one task was supported at a time. The authors divided the entire task into different parts and transmitted them simultaneously. To be more specific, if the goal was to transmit a single image, the authors divided the image into four equal segments and transmitted them at the same time. Additionally, with careful design of RISs, it is possible to support multiple tasks simultaneously, which is not achievable with AI on digital hardware. As shown in Fig.~\ref{background}(d), the authors in~\cite{Meta2022LuoLight} performed object recognition using vertically polarized waves and hand-written number recognition using horizontally polarized waves. Even though the signals with different tasks passed through the same RIS simultaneously, different transformations can be achieved if these signals have orthogonal polarizations. Therefore, this not only increases system capacity, but also enables support for diverse services.
\item \textbf{Low computation power consumption:} The power consumption required in the proposed scheme can be divided into two parts: the energy cost by the neurons and the energy cost by the controlling system. The energy consumed by each neuron depends on the transmission frequency and the manufacturing technique. Generally, each neuron requires very low power. The energy of the controlling system is determined by the type of FPGA used. In \cite{Meta2022CuiNature}, four layers of RISs were used to implement DNN functions on wireless signals, with an average power consumption of about 100 Watts for the entire system. In comparison, performing the same task using digital hardware (such as Tesla V100 or A100) consumes an average power of 400 Watts. The authors also mentioned that the computing efficiency of the RIS-based DNN can be further improved by using a small power amplifier and operating at a lower energy level. Additionally, for a large-scale network, the increased power required by RIS-based AI is much smaller compared to hardware-based AI. This further highlights the attractiveness of the proposed scheme in handling complex tasks.
\item \textbf{Work in a wide spectral range:} Like hardware-based AI, the transmission frequency should also be considered in designing the structure of RISs, as one RIS cannot achieve the expected programable performance in all frequency bands. However, thanks to the rapid development of RISs, an enough wide spectral range can be supported and the performance of the proposed scheme is satisfactory to meet the requirements. For example, the working band of the RIS ranges from 5.35 to 5.7 GHz, and the element can realize the transmission energy reduction down to -22 dB and the amplification up to 13 dB, which offers a enough wide range for weight matrix~\cite{Meta2022CuiNature}.
\item\textbf{High security:} By separating the data and control flows at the hardware level, we can enhance security measures.
\item \textbf{Real-time programmable capability:} The real-time programmable feature of the initial RIS layer guarantees that the conversion of information data from serial to parallel and the modulation process can be performed instantly, thereby reducing the signal processing time.
\item \textbf{Benefits brought by CvCNN:} The suggested approach is based on CvCNN. CvCNN has demonstrated superior performance in certain tasks involving image and voice signals, and it is reasonable to expect similar benefits when applying it to semantic communications.
\item \textbf{Potentials in other wireless communication applications:} The functions of an RIS layer can be seen as a type of transformation. Research has demonstrated that RIS can perform various types of transformations, including Fourier and Laplace transforms of spatial signals~\cite{Meta2020DingPhono, Pan2021LaplacePhoto}, as well as convolution of sequence signals~\cite{Sanchez2022Arxiv}. Therefore, in addition to semantic communications, the proposed scheme has significant potential to be used for signal processing in various wireless communication applications in the future.
\end{itemize}

However, it is important to acknowledge that there are limitations to the proposed scheme, primarily due to the constraints posed by the current manufacturing technique of RISs.
\begin{itemize}
\item \textbf{Requirement in consecutive adjustment:} To fully harvest the performance benefits of a CNN, it is expected that the parameters of each neuron should be adjusted consecutively within a feasible range. Although there are already manufactured consecutively phase-adjusted RISs, currently widely used programmable RISs can only support a limited number of phases. To mitigate the impact of non-continuous phase adjustments, the utilization of discrete optimization algorithms~\cite{Meta2022CuiNature} or digital pre-coding~\cite{An2023Arxiv} may be feasible solutions.
\item \textbf{Unstable nonlinear transformation:} In general, it is unstable to perform nonlinear transformation on programmable RISs, and thus the errors between simulation and experimental results are inevitable, which will affect the performances of the proposed scheme in practical applications.
\item \textbf{Additional transmission power loss in RISs:} As semantic encoding occurs after modulation, the transmit power refers to the emission power from the final RIS layer. Therefore, it is important to consider the power loss caused by signals passing through RISs. The power loss of a signal transmitted or reflected by a single RIS is influenced by various factors, including polarization, frequency, and manufacturing technique~\cite{Wang2020PowerPRA}. However, due to space limitations, a detailed elaboration on these factors is beyond the scope of this paper.
\item\textbf{Restrictions by RIS computation capability:} It has been demonstrated that RISs can achieve the fundamental functionalities of DNN and CNN. However, DNN and CNN may not fulfill the requirements of every task. Additional research is needed to investigate the potential of other AI algorithms in RISs.
\end{itemize}

In conclusion, the unresolved concerns regarding the proposed schemes primarily stem from limitations in existing manufacturing techniques and a lack of sufficient research on RIS-based semantic communications. However, with the progress being made in RISs and related fields, we believe that these issues will be addressed in the coming years.

\vspace{0.25in}
\section{CONCLUSIONS}
A new way of communicating using RISs was proposed to enable wireless-based AI. In this scheme, a RIS layer acts like a DNN layer, and a mete-atom functions as a neuron. The process of wireless signals passing through a RIS layer is similar to how a DNN layer operates. As a result, this RIS-based semantic communication approach offers the same functionalities as traditional hardware-based semantic communications, but with several advantages such as fast computation, support for multiple tasks at once, low power consumption, broad spectral range, high security, real-time programmability, etc. To demonstrate the effectiveness of the proposed scheme, an example of data-reconstruction-oriented semantic communications was provided, and simulation results showed improved PSNR compared to traditional schemes. Additionally, we have identified the challenges in RIS-based semantic communications as areas for future work.


\vspace{0.25in}
\balance

\vfill

\end{document}